\documentclass[10pt]{iopart}
\pdfoutput=1
\usepackage{graphicx}

\usepackage{gensymb}
\usepackage[breaklinks=true,colorlinks=true,linkcolor=blue,urlcolor=blue,citecolor=blue]{hyperref}
\usepackage{multirow}
\usepackage{array}
\begin{document}

\title[Formation of a boron-oxide termination for the (100) diamond surface]{Formation of a boron-oxide termination for the (100) diamond surface}

\author{Alex K. Schenk$^{1}$, Rebecca Griffin$^{1}$, Anton Tadich$^{1,2}$, Daniel Roberts$^{3}$, Alastair Stacey$^{3,4}$}

\address{$^{1}$ Department of Mathematical and Physical Sciences, School of Computing, Engineering and Mathematical Sciences, La Trobe University, Victoria 3086, Australia}
\address{$^{2}$ Australian Nuclear Science and Technology Organisation (ANSTO), Clayton, VIC 3168, Australia}
\address{$^{3}$ School of Science, RMIT University, Melbourne, Vic 3001, Australia}
\address{$^{4}$  Princeton Plasma Physics Laboratory, Princeton University, 100 Stellarator Rd, Princeton, NJ, 08540, USA}

\ead{A.Schenk@latrobe.edu.au}

\vspace{10pt}

\begin{abstract}
A boron-oxide termination of the diamond (100) surface has been formed by depositing molecular boron oxide $\rm{B_2O_3}$ onto the hydrogen-terminated (100) diamond surface under ultrahigh vacuum conditions and annealing to 950\celsius. The resulting termination was highly oriented and chemically homogeneous, although further optimisation is required to increase the surface coverage beyond the 0.4~ML achieved here. This work demonstrates the possibility of using molecular deposition under ultrahigh vacuum conditions for complex surface engineering of the diamond surface, and may be a first step in an alternative approach to fabricating boron doped delta layers in diamond. 
\end{abstract}

\ioptwocol

\section{Introduction}
Engineering (100) diamond surfaces using chemical modification has long been explored for diverse applications including quantum electronic \cite{crawford2021surface} and spintronic devices \cite{xing2020strong,edmonds2015spin}, cold cathode electron emitters \cite{james2021review} and high power, high frequency electronic components \cite{pakes2014diamond,denisenko2005diamond}. As diamond is considered relatively chemically inert \cite{szunerits2008different}, early work focussed on surface terminations which could be efficiently produced using highly reactive acids and aggressive plasma conditions, limiting the pool of high quality terminations to monoatomic species - namely hydrogen \cite{ley2011preparation}, oxygen \cite{li2021systematic}, nitrogen \cite{stacey2015nitrogen} and halogen-terminated diamond surfaces \cite{rietwyk2013work}. In more recent years the development of ultrahigh vacuum (UHV) based surface functionalisation processes, in which reactive conditions can be produced that would otherwise not be feasible, have enabled termination of diamond by alkali metals \cite{o2013diamond,o2015extremely} and group IV elements \cite{schenk2015formation,sear2017germanium}, with the latter case allowing for further chemical modification of the terminated surface using oxygen \cite{schenk2016oxidation, sear2018thermal} and fluorine \cite{schenk2022fluorination}. Such demonstrations highlight the potential of UHV-based functionalisation methods to enable new approaches to engineering the properties of diamond.

Boron is a common p-type dopant added to otherwise insulating diamond in order to create controlled conductivity, typically being added either during chemical vapour deposition (CVD) growth of diamond or following growth with ion implantation \cite{kalish1999doping}. Both approaches work well in producing tunable bulk conductivity, with changes in boron concentration allowing diamond to be tuned from the natural insulating state to a semiconducting \cite{chrenko1973boron} or superconducting \cite{yokoya2005origin} state. Over the last 30 years there have been many attempts at producing boron doped $\delta$-layers - highly doped layers only a few atomic layers in thickness - in otherwise insulating diamond with both ion implantation \cite{uzan1998formation} and CVD \cite{vikharev2016novel,el2008characteristics,butler2017nanometric}, with the goal of forming conductive diamond layers with quantum enhanced electrical transport and quantised electronic states \cite{butler2017nanometric,chicot2014electronic,fiori2012situ}. While $\delta$-layers as thin as 1.2~nm (approximately 13 atomic layers) have been fabricated in diamond \cite{butler2017nanometric}, experimental characterisation of the transport \cite{chicot2014electronic} and electronic properties \cite{pakpour2020occupied} of boron-doped diamond $\delta$-layers has failed to demonstrate the extremely high carrier mobility predicted for perfectly delocalised carriers in nominal $\delta$-layers \cite{chicot2014electronic,butler2017nanometric}. This is possibly due to the difficulty in producing a sufficiently sharp transition from doped to undoped diamond \cite{pakpour2020occupied}, or a need to produce even narrower $\delta$-layers for these effects to be observable \cite{chicot2014electronic}, two challenges which are difficult to overcome using CVD or ion implantation. Additionally, while there is considerable theoretical \cite{shen2020monolayer,zhaolong2022tunable,sun2022boron} and experimental \cite{chen2023interface,yang2020epitaxial,koizumi1990epitaxial} work exploring heterojunctions between diamond and boron-based materials - notably boron nitride - for electronic devices, CVD and ion implantation are poorly suited to forming high quality boron-containing layers at the diamond surface. Taking inspiration from molecular beam epitaxy interface engineering \cite{ploog1987molecular} and $\delta$-layer growth in phosphorus-doped silicon systems \cite{zhang2021advances}, we believe that developing the capability to chemically attach boron compounds to the diamond surface in an ordered and controlled fashion may be a key step to addressing the challenges in $\delta$-doping of diamond and forming interfaces between diamond and boron-containing materials. 

This article demonstrates a UHV-based method for forming a boron-oxide surface termination for (100) diamond using $\rm{B_2O_3}$ as a molecular precursor. X-ray Photoelectron Spectroscopy (XPS) and Near Edge X-ray Absorption Fine Structure (NEXAFS) spectroscopy analysis indicates that the surface is only partially terminated with boron-oxide, with an ultimate coverage of 0.4~ML and the remaining surface sites being bare diamond dimers, pointing to a need for process optimisation. Despite this, the termination is largely chemically homogeneous and well oriented, both promising characteristics for surface-based chemical synthesis and forming high quality layers. Having such an oriented and well-defined boron-containing layer at the surface may be a first step to an alternative surface termination and subsequent diamond overgrowth approach to forming boron doped diamond delta layers \cite{kuntumalla2021nitrogen}. Furthermore, our success in bonding boron-oxide to the surface suggests it may be possible to bond other boron-based molecular species to diamond in an ordered fashion using a UHV-based functionalisation approach, thereby opening possibilities for developing more complex UHV-based synthesis processes for boron-based heterostructure engineering at the diamond surface. 

\section{Experimental Section}
This work has been performed on a $4~\rm{mm} \times 4~\rm{mm}$ CVD-grown type-IIa single crystal (100) oriented substrate (Element Six), onto which a (100) oriented boron doped overlayer (B concentration in the range $1\times10^{18}$ -- $1\times10^{19}$ B/cm\textsuperscript{-3}) was grown at the Melbourne Centre for Nanofabrication (MCN), to prevent charging in the course of measurements. Following the overlayer growth, the sample was hydrogen-terminated in a microwave hydrogen plasma containing a small amount of methane and trimethylborane (TMB) for approximately 5 minutes at 80 Torr, with the sample maintained at a temperature of approximately 850\celsius\, throughout that period. At the end of 5 minutes, the methane and TMB gas flows are stopped and the microwave power turned down and then off. The inclusion of the methane and TMB during the hydrogen-termination process effectively acts to grow a thin layer of doped diamond during the termination, ensuring that the surface remains conductive despite the etching nature of the hydrogen plasma.

All subsequent sample preparation steps and measurements were conducted at the Soft X-ray Spectroscopy beamline at the Australian Synchrotron. The beamline endstation is equipped with a SPECS Phoibos 150 hemispherical analyser for performing high resolution core-level photoelectron spectroscopy (XPS) and a retarding grid analyser with channeltron detector for partial electron yield (PEY) NEXAFS. The sample was mounted inside a Ta envelope, leaving most of the top face of the sample uncovered, on a sample holder with an underlying electron beam heater. A K-type thermocouple in direct contact with the sample envelope and an optical pyrometer were simultaneously used to monitor the sample temperature during annealing steps. Following introduction to the UHV endstation, the hydrogen-terminated substrate was annealed at 450\celsius\, for 1 hour to remove atmospheric adsorbates \cite{schenk2016high}. Boric anhydride ($\rm{B_2O_3}$), sourced from Sigma-Aldrich, was then deposited by thermal evaporation using a Knudsen cell at a source temperature of 900\celsius, to a coverage of 2.5~ML as estimated by photoelectron attenuation measurements acquired with a photon energy of 400~eV; the details of these calculations are presented in the Supplementary Information. Finally, the sample was annealed to 950\celsius\, for 30 minutes, removing the hydrogen from the diamond surface \cite{schenk2016high, graupner1998high,graupner1999surface} and allowing bonding between the $\rm{B_2O_3}$ and the underlying diamond substrate. Between each preparation step the sample was characterised using both XPS and NEXAFS. 

XPS measurements of the C1s, B1s and O1s core levels were performed using photon energies of 350~eV, 225~eV and 600~eV respectively in order to maximise surface sensitivity, while survey scans were acquired using a photon energy of 850~eV. The binding energy (BE) scale of all spectra are referenced to the system Fermi level by measuring the $\rm{Au4f_{7/2}}$ core level of a gold foil in electrical contact with the sample and setting the binding energy to 84.00~eV. Core level spectra were fitted by first applying a Shirley background subtraction \cite{shirley1972high} and then using symmetric Voigt components with Lorentzian widths of 0.15~eV for the C1s core level \cite{schenk2016high, graupner1998high} and 0.20~eV for the O1s core level \cite{schenk2016oxidation,roodenko2007time}. We are unable to find a consistent value in the literature for the B1s core level Lorentzian width, and so performed our analysis of the B1s core level with Lorentzian width values in the range 0.07 - 0.15~eV to determine a value which was was consistently suitable for our spectra. We find that a Lorentzian width of 0.12~eV is most appropriate for our data, as determined from the residuals of the fits, and so this is what is used for the fits presented in this paper. However, we note that using other values in the 0.07 - 0.15~eV range does not change the qualitative results of our analysis, and only changes our quantitative results by an amount which is within the error of the analysis. 

PEY NEXAFS measurements were conducted using linearly polarised soft x-rays generated by the beamline undulator light source. Energy alignment of the spectra was achieved by measuring stable, calibrated reference foils in parallel with each measurement. Each spectra is normalised using the beamline intensity acquired during measurement. 

Bare (100) diamond reference data presented in Figure~\ref{fig:DimerProof} has been acquired from a sample prepared using the same hydrogen termination process described above. This sample was then loaded into vacuum, annealed to 450\celsius\, for 1 hour to remove atmospheric adsorbates, and subsequently annealed at 950\celsius\, for 30 minutes to remove all hydrogen \cite{schenk2016high,graupner1998high,graupner1999surface}. 

\section{Results and Discussion}
\begin{figure*}
	\centering
	\includegraphics{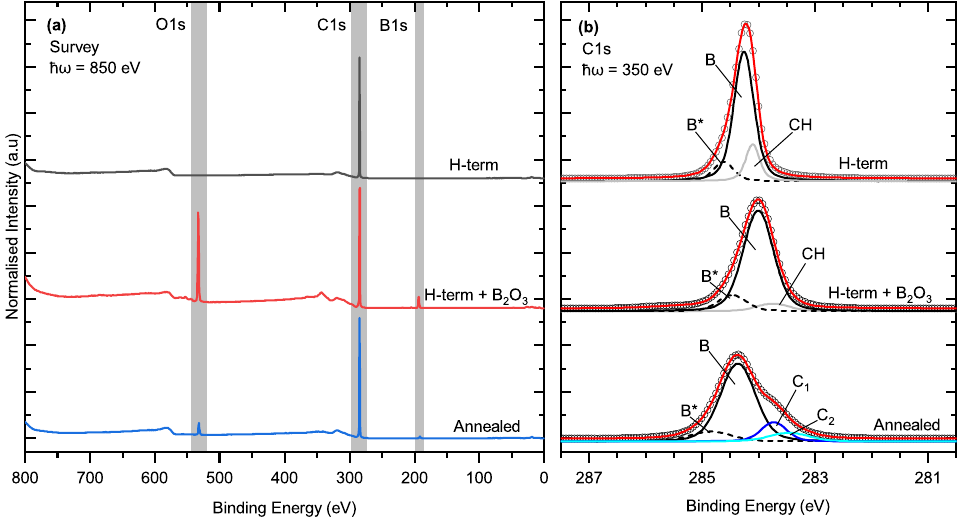}
	\caption{X-ray photoelectron spectroscopy (a) survey scans ($\hbar\omega = 850$~eV) and (b) high resolution C1s core level scans ($\hbar\omega = 350$~eV) of the sample at each stage of the preparation. Fitting components in (b) are vertically offset for clarity.}
	\label{fig:CarbonSpec}
\end{figure*}

Figure~\ref{fig:CarbonSpec} shows the XPS survey scans and surface sensitive, high resolution C1s XPS acquired at each stage of this experiment following the initial anneal to clean the sample of atmospheric contaminants. The spectroscopy from the hydrogen-terminated (100) diamond surface is consistent with the literature \cite{schenk2016high}. The C1s core level for the hydrogen-terminated surface has been fit using a 3 component model established in previous work \cite{schenk2016high}, in which the components $B$ and $B^{*}$ are associated with the diamond bulk, and $CH$ is associated with the carbon atoms at the surface participating in $\rm{C-H}$ bonds. 

Following the $\rm{B_2O_3}$ deposition, the survey scan shows clear peaks in the B1s and O1s regions, with no other elements being added to the surface. The C1s core level at this stage requires no new components to fit - there is a slight increase in the width of the components, explained by an increase in inelastic scattering as a result of the presence of an overlayer, and a slight shift of the entire core level by $-0.13\pm0.05$~eV, indicating some slight p-type surface transfer doping of the diamond surface by the $\rm{B_2O_3}$ overlayer, consistent with observations in other metal-oxide layers on the hydrogen-terminated diamond surface \cite{crawford2021surface}. Combined, the observations at this stage of the experiment indicate the presence of both boron and oxygen on the surface, and that the hydrogen termination passivates the surface against any chemical reaction with the $\rm{B_2O_3}$ film, similar to other experiments involving UHV functionalisation of diamond \cite{schenk2015formation,sear2017germanium,schenk2020development}. 

A more dramatic change follows the 950\celsius\, anneal. In the survey scan both the B1s and O1s core levels are still present, albeit decreased in intensity relative to the C1s core level, suggesting the loss of both oxygen and boron with annealing; photoelectron attenuation calculations (detailed in the Supplementary Information) indicate that the surface coverage of boron is 0.4~ML following the anneal. This reduction is not unreasonable given that the $\rm{B_2O_3}$ evaporation temperature (900\celsius) is lower than the hydrogen desorption temperature (950\celsius\, \cite{graupner1998high}), and suggests that any observed chemical changes at the surface will be the resulting of competing reaction and desorption processes; we will return to this point later. The C1s core level following the anneal requires four components to fit. Two of these are the original diamond bulk $B$ and associated $B^{*}$ component, the latter of which - in keeping with the subsurface carbon assignment made in previous work \cite{schenk2016high} - we have fixed in area at 15\% of the $B$ component, but allowed to shift in binding energy relative to the bulk peak. This leaves two components, labelled as $\rm{C_1}$ and $\rm{C_2}$ in Fig.~\ref{fig:CarbonSpec}(b), with chemical shifts relative to the diamond bulk (B) of $-0.63\pm0.05$~eV and $-0.90\pm0.05$~eV respectively, which we attribute to surface-related species; these components present in a 1.52(3):1 ($\rm{C_1}$:$\rm{C_2}$) intensity ratio. We note that there are no new components to higher binding energy of the diamond bulk, which would be present if there were carbon-oxygen bonding at the surface \cite{maier2001electron,dontschuk2023x}.We can therefore infer that these components represent either carbon-boron or carbon-carbon bonding at the surface. This is based on the logic that, to first order, the core level shift produced by bonding between two elements is largely dictated by the difference in the Pauling electronegativity of the two elements and the number of bonds between them \cite{gao2006solid}. For the case of carbon and boron - Pauling electronegativities of 2.55 and 2.05 respectively - bonding at the surface, we would therefore expect the corresponding C1s components to have a core level shift to lower binding energy of the diamond bulk component. However, it is possible that the boron and oxygen in the survey scan is simply an indication that some amount of $\rm{B_2O_3}$ remains physisorbed after the anneal, and so we are not able \textit{a priori} to say more about whether these components are carbon-boron or carbon-carbon related without first examining the adlayer-related chemistry in greater detail.  

\begin{figure}
	\centering
	\includegraphics{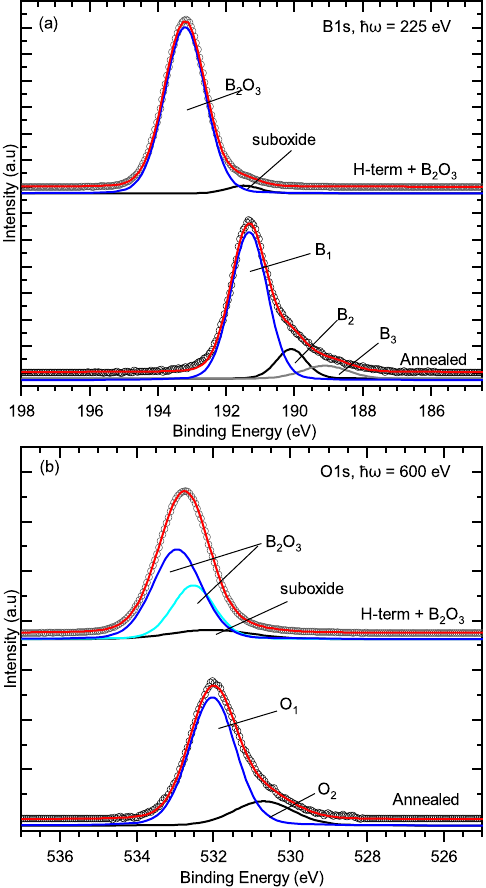}
	\caption{High resolution photoelectron spectra of (a) boron (B1s, $\hbar\omega = 225$~eV) and (b) oxygen (O1s, $\hbar\omega = 600$~eV) after $\rm{B_2O_3}$ deposition and following the 950\celsius\, anneal. Components are vertically offset for clarity.}
	\label{fig:AdlayerXPS}
\end{figure}

Figure~\ref{fig:AdlayerXPS} shows the B1s and O1s core level after deposition and after the 950\celsius\, anneal. When analysing the post-deposition XPS spectra, we have worked from the knowledge that molecular $\rm{B_2O_3}$ contains two boron atoms in the same $\rm{O-B=O}$ chemical environment and three oxygen atoms, two of which are in a $\rm{O=B}$ environment and the third which is in a $\rm{B-O-B}$ environment. This means, for a unreacted and stoichiometrically correct $\rm{B_2O_3}$ adlayer, we would expect to see one component in the B1s core level, and two components with a 2:1 intensity ratio in the O1s core level. Our fit of the post-deposition B1s core level in Fig.~\ref{fig:AdlayerXPS}(a) contains two components, with the higher binding energy component ($\rm{BE_{B1s}} = 193.3\pm 0.05~\rm{eV}$) being more than 96\% of the total core level intensity. In the O1s spectra, Fig.~\ref{fig:AdlayerXPS}(b), we find three components are required for fitting, with the larger two components having a 1.9:1 area ratio and these two components combined representing 91\% of the core level intensity. We therefore ascribe the largest component in the B1s and the two largest components in the O1s core level to $\rm{B_2O_3}$, and that the minority components in these spectra are likely the result of a small amount of boron suboxide \cite{music2002synthesis,wang1992characterization,ong2004x}, such as $\rm{B_2O_2}$ or $\rm{B-O}$, which would be expected to have components to lower binding energy of the $\rm{B_2O_3}$ components. This suboxide is likely produced due to some amount of thermal decomposition of $\rm{B_2O_3}$ during deposition, or beam-induced decomposition of the molecule under synchrotron irradiation \cite{edmonds2012surface}; given the relatively small amount of suboxide, it is reasonable for us to assume it will not significantly influence the results of our experiment. 

\begin{table}
	\caption{Summary of B1s and O1s core level component (comp.) binding energies (BE) and relative intensities for the post-anneal surface. Fits shown in Fig.~\ref{fig:AdlayerXPS}.}
	\label{tab:XPSFit}
	\centering
	\begin{tabular}{|c|c|c|c|}	
		\hline
		Core level & Comp. & BE (eV) & Intensity (\%) \\
		\hline
		\multirow{3}{*}{B1s} & $\rm{B_1}$ & $191.43\pm0.05$ & $79\pm2$ \\ \cline{2-4}	
		& $\rm{B_2}$ & $190.20\pm0.05$ & $13\pm2$ \\ \cline{2-4}
		& $\rm{B_3}$ & $189.21\pm0.05$ & $8\pm2$ \\ \hline
		\multirow{2}{*}{O1s} & $\rm{O_1}$ & $532.03\pm0.05$ & $81\pm2$ \\ \cline{2-4}
		& $\rm{O_2}$ & $530.70\pm0.05$ & $19\pm2$ \\ \hline
	\end{tabular}	
\end{table}

\begin{figure*}
	\centering
	\includegraphics[width = 170mm]{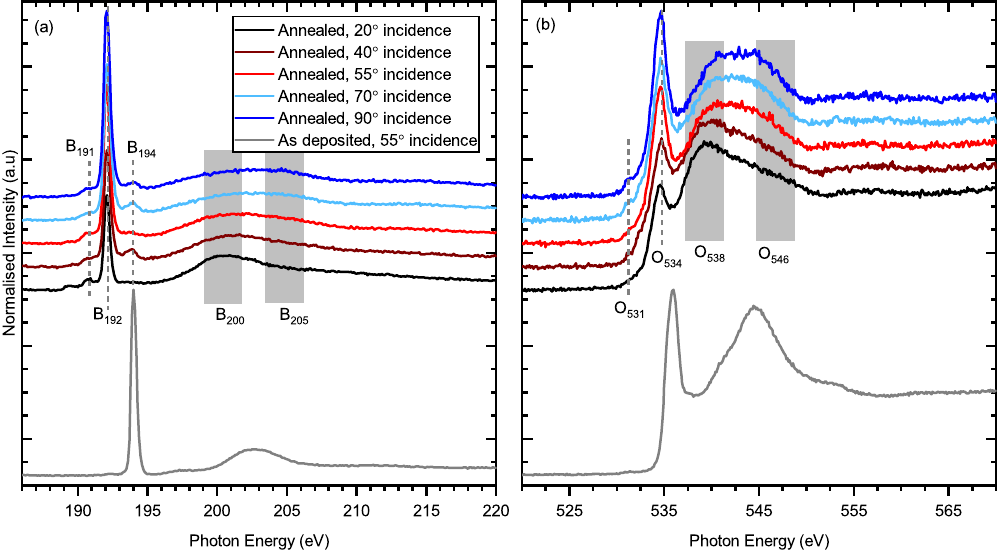}
	\caption{NEXAFS acquired at the (a) boron and (b) oxygen K-edges. Each panel shows the data acquired after the 950\celsius\, anneal at 20\degree, 40\degree, 55\degree, 70\degree\, and 90\degree\, beam incidence, as well as post-deposition NEXAFS acquired at 55\degree\, incidence for comparison. The legend in (a) applies for both panels.}
	\label{fig:AdlayerNEXAFS}
\end{figure*}

Following the 950\celsius\, anneal we see distinct changes in both the O1s and B1s core levels; Table~\ref{tab:XPSFit} contains details of the fits following the anneal. The B1s core level now requires three components to fit - $\rm{B_1}$, $\rm{B_2}$ and $\rm{B_3}$ - and the entire core level is shifted to lower binding energy with respect to the as-deposited $\rm{B_2O_3}$ B1s core level. This suggests that the boron chemistry has changed from the $\rm{O-B=O}$ chemical environment in $\rm{B_2O_3}$. Given the observation of new surface-related components in the C1s core level following the 950\celsius\, anneal, and that the valence of boron limits it to forming 3 bonds - thereby preventing higher amounts of oxidation than that present in $\rm{B_2O_3}$ - we infer from this change that at least one boron-carbon bond has replaced a boron-oxygen bond. In the O1s core level there are now two new components, $\rm{O_1}$ and $\rm{O_2}$, clearly indicating that oxygen still remains at the surface following the anneal in a chemically altered form. Given the lack of any $\rm{C-O}$ component in the C1s core level (Fig.~\ref{fig:CarbonSpec}(b)), the only explanation for the presence of oxygen following the anneal is that there is still boron-oxygen bonding at the surface. Additionally, given that the B1s core level analysis indicates the bonding of boron to the diamond surface, it is clear that the termination which has formed involves boron-oxygen moieties, bonding through the boron atom. We note that, based on the component intensities shown in Table~\ref{tab:XPSFit}, it appears that the $\rm{B_1}$ and $\rm{O_1}$ components are likely related, given their similar intensities, and if this is the case then the chemical environment giving rise to these components makes up approximately 80\% of the boron-oxide related surface termination. Given that the diamond (100) surface is known to have a high step edge and defect density, in which approximately 15\% of all surface sites \cite{schenk2016high} are not complete surface dimers, the minority components ($\rm{B_2}$, $\rm{B_3}$, $\rm{O_2}$) in the core levels may be the result of boron binding into step edges and defect sites and thereby adopting alternative chemistries to the majority boron-oxygen moiety. We are unable to determine from XPS alone the bonding characteristics of this majority species, for instance the bonding order between the boron and surrounding oxygen and carbon and whether the termination is highly oriented, and so turn to angle-dependent NEXAFS measurements of the termination to gain more information about the surface termination. 

Figure~\ref{fig:AdlayerNEXAFS} shows NEXAFS acquired at five angles (20\degree, 40\degree, 55\degree, 70\degree\, and 90\degree) of beam incidence with respect to the surface plane at the boron and oxygen K-edges following the 950\celsius\, anneal. For comparison, the 55\degree\, NEXAFS acquired on the hydrogen-terminated diamond surface following $\rm{B_2O_3}$ deposition is also provided. Given the difficulty in establishing a unique fitting model for NEXAFS spectra, which may contain multiple overlapping components, for a surface which is not yet well understood, we elect to explore this data in a qualitative fashion at the present time rather than assign meaning to an ambiguous fitting model. This approach requires us to assume that the species creating the majority component in the XPS ($\rm{B_1}$, $\rm{O_1}$) will be the primary contribution to the corresponding NEXAFS spectra, and so the dominant features in the NEXAFS spectra are the main focus of our evaluation. 

What we clearly observe from the NEXAFS measurements is that there has been a change in the chemical structure of the boron and oxygen, given that the spectra have changed with annealing, confirming our assessment from the XPS measurements presented above. Additionally, both the boron and oxygen K-edge spectra have multiple features where the intensity varies as a function of incidence angle; features and regions of interest are labelled in Figure~\ref{fig:AdlayerNEXAFS}. An observed angular dependence in the normalised intensity of NEXAFS spectra is common for measurements performed on surfaces with a highly oriented termination \cite{nefedov2013advanced}, as the intensity variation is a result of polarisation selection rules which contain, amongst other things, the relative orientation of the electric field vector of the x-rays and the real space orientation of the unoccupied molecular orbitals at the surface \cite{Stohr,hahner2006near}. Therefore, the observation of angle-dependent behaviour in features in NEXAFS data is an indicator of a high degree of spatial ordering of bonds related to those features. As sharp low energy peaks in NEXAFS spectra are typically attributed to $\pi^*$ resonances \cite{Stohr,hahner2006near} - indicating the existence of a double or triple bond - the sharp peaks in the boron ($\rm{B_{192}}$) and oxygen ($\rm{O_{534}}$) K-edges, with the same angular dependence, suggest the existence of a $\rm{B=O}$ surface termination. Additionally, the intensity maximum of this $\pi^{*}$ resonance being at 90\degree\, incidence suggests that this is double bond points normal to the surface. Such a $\pi^{*}$ resonance should be accompanied by a $\sigma^{*}$ resonance with opposing angular variation which would typically appear to higher energy \cite{Stohr,hahner2006near,nefedov2013advanced}. Certainly there are broad regions with angle-dependent intensity in both the boron and oxygen NEXAFS to higher energy of this $\pi^{*}$ resonance ($\rm{B_{200}}$, $\rm{B_{205}}$, $\rm{O_{538}}$, $\rm{O_{546}}$), and two of these regions ($\rm{B_{200}}$, $\rm{O_{538}}$) have an intensity maximum at 20\degree, appropriate for a $\sigma^{*}$ transition associated with a $\rm{B=O}$ bond normal to the surface. However, we are unable to definitively make this assignment given that there is no clear peak that can be identified in either spectra, possibly a combined result of $\sigma^{*}$ transitions often being broad transitions, and there being multiple overlapping components in these regions of the spectra. Indeed, given that the XPS (Fig.~\ref{fig:AdlayerXPS}) showed the existence of minority species at the surface, we can expect that there will be features in the NEXAFS spectra which are generated by these additional chemical moieties. This offers an explanation for the distinct low intensity features ($\rm{B_{191}}$, $\rm{B_{194}}$, $\rm{O_{531}}$), but we cannot rule out that there are additional features within the $\rm{B_{200}}$, $\rm{B_{205}}$, $\rm{O_{538}}$ and $\rm{O_{546}}$ regions that are also associated with such species. 

\begin{figure}
	\centering
	\includegraphics{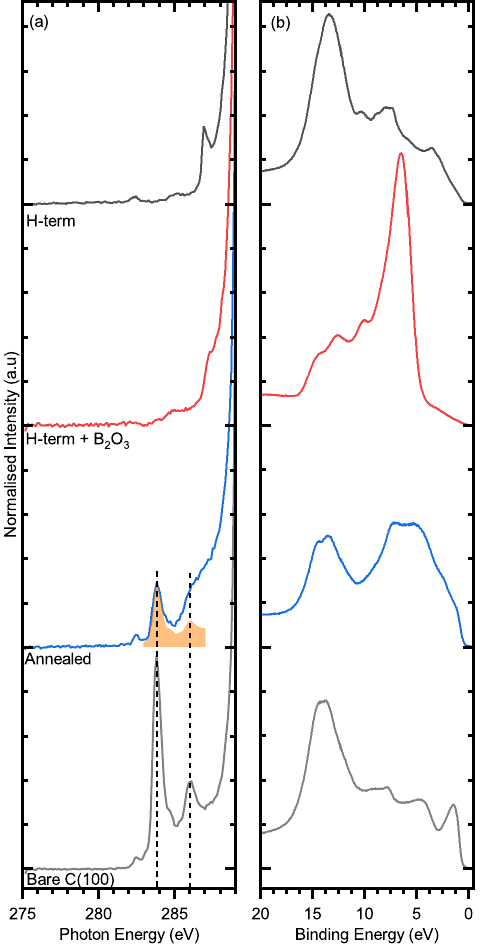}
	\caption{The pre-edge region of carbon K-edge NEXAFS acquired at 55\degree\, beam incidence (a) and valence band spectra (b) acquired at each stage of this experiment, alongside reference spectra for a bare (thermally depassivated) diamond (100) surface. The filled area overlaid on the ``Annealed'' spectra indicates the area accounted for by the presence of bare dimers, as discussed in text. Dashed lines indicate features discussed in text.}
	\label{fig:DimerProof}
\end{figure}

Having established some understanding of the likely chemical bonding of each of the elements at the surface based on our XPS analysis, we now turn to the identification of the two surface-related carbon species. Our observations from the NEXAFS spectra suggest that the surface boron is primarily in a $\rm{B=O}$ environment; given that boron is typically restricted to forming 3 bonds, this would imply that the majority of the boron-bonded surface sites are in a $\rm{C-B=O}$ environment. As noted above, the survey scan (Fig.~\ref{fig:CarbonSpec}(a)) indicates the loss of boron and oxygen following the 950\celsius\, anneal, and photoelectron attenuation calculations suggest that the boron coverage following the anneal is 0.4~ML or approximately two boron atoms per five surface carbon atoms. Reconciling this with our assumption for the termination chemistry for a complete surface would imply that two of these five carbon atoms form a $\rm{C-B=O}$ structure, forming one C1s XPS component, while the remaining carbon atoms adopt an alternative carbon-carbon bonding arrangement, producing a second component in the C1s. By itself this is not implausible for a complete monolayer termination of the surface, although would require an unusual surface reconstruction that has not previously been observed for diamond. However, we must also consider the possibility of a partial surface termination, given that the chemical shift of component $\rm{C_1}$ relative to the diamond bulk ($-0.63\pm0.05$~eV) is similar to the chemical shift of the C(100) reconstructed surface dimer \cite{schenk2016high}. Figure~\ref{fig:DimerProof} shows the 55\degree\, incidence pre-edge carbon K-edge NEXAFS and valence band spectra for each stage of this experiment, alongside reference data acquired on a thermally depassivated (100) diamond surface. The feature at 283.8~eV in the annealed spectra lines up well with the strong feature in the thermally depassivated pre-edge which is attributed to the $\pi^*$ resonance of the $\rm{C=C}$ surface dimer of bare diamond \cite{shpilman2014near}. These dimers also produce the 286~eV $\sigma^{*}$ feature  in the thermally depassivated pre-edge \cite{shpilman2014near}, where the annealed spectra from this experiment also has some increased intensity, although there is not a distinct 286~eV feature. As the $\pi^{*}$ and $\sigma^{*}$ feature should appear in approximately the same proportion for any density of dimers \cite{Stohr,hahner2006near}, we have scaled the 283~eV - 287~eV region of the bare diamond spectra such that the $\pi^{*}$ feature matches the size of the $\pi^{*}$ feature in the annealed spectra, in order to estimate the size of the $\sigma^{*}$ feature. This estimate overlays the annealed spectra (shown as a shaded orange region) in Fig.~\ref{fig:DimerProof}(a), clearly demonstrating that surface dimers alone are insufficient to explain the pre-edge structure. If we turn to the valence band spectra (Fig.~\ref{fig:DimerProof}(b)), we likewise see that while there are some qualitative similarities between the annealed and bare diamond spectra, and even between the annealed and $\rm{B_2O_3}$-deposited spectra, the annealed spectra is distinct from those others, a clear indication that this process gives rise to a surface electronic structure which is unique from the bare C(100) and $\rm{B_2O_3}$ deposited surfaces. Keeping this information in mind and returning to our core level analysis, we infer that the $\rm{C_1}$ component in the C1s core level is related to $\rm{C=C}$ surface dimers that have formed during the annealing process, and attribute the $\rm{C_2}$ component to carbon atoms participating in carbon-boron bonds. The chemical shift of the $\rm{C_2}$ component is not trivial to interpret in terms of how many carbon-boron bonds each carbon atom is participating in given the few examples in the literature of low electronegativity elements bonding directly to the diamond surface, which would otherwise provide guidance on the anticipated magnitude of the shift. However, the observed intensity ratio between the $\rm{C_1}$ and $\rm{C_2}$ components (1.52:1) is appropriate for a surface where 60\% of the surface sites are bonded in a bare dimer configuration ($\rm{C_1}$) and 40\% are forming carbon-boron bonds ($\rm{C_2}$), a situation which can be reconciled with the calculated 0.4~ML coverage only if every carbon atom in the $\rm{C_2}$ environment is participating in a single carbon-boron bond, suggesting that this is likely the case. 

Finally, we consider future directions for this work. While we are unable to unambiguously assign the complete structure of our termination based on the data acquired in this study, what is clear from our work is that depositing $\rm{B_2O_3}$ onto hydrogen-terminated (100) diamond and subsequently annealing to 950\celsius\, produces a two-element surface termination in which a boron-oxygen moiety is bonded to the diamond (100) surface through the boron atom, and approximately 80\% of the bonded surface sites are identical and highly oriented. Our NEXAFS analysis strongly suggests that it is a $\rm{B=O}$ moiety, which would be consistent with published work exploring thermal decomposition of $\rm{B_2O_3}$ for surface boron incorporation in silicon (100), where the $\rm{B=O}$ bond remains \cite{korobtsov2006b2o3} following the initial decomposition, before being broken by other processes. Establishing a refined NEXAFS model, which will enable the extraction of further details surrounding the minority species and determination of the exact geometry of the surface termination, will require more significant experimental and theoretical work.

This approach to forming a two-element surface termination for diamond is distinct from other demonstrated termination processes, which have been realised using a two-stage process and are typically less ordered \cite{o2013diamond,o2015extremely,schenk2022fluorination}, and demonstrates that ordered, more complex terminations may be achievable on diamond through an appropriate choice of molecular precursor. The high degree of chemical homogeneity and orientation may be useful in developing reaction schemes where the oxygen is subsequently replaced by some other chemical moiety, enabling surface-based synthesis of boron phases at the diamond surface. That we were only able to achieve a surface coverage of 0.4~ML may be due to steric limitations at the surface, as is the case for germanium-terminated diamond \cite{sear2017germanium}, but given the small size of boron and the demonstrated ability to form full monolayer terminations of diamond with similarly sized metal oxides \cite{o2013diamond,o2015extremely}, it is more likely that this coverage is a result of competing desorption and reaction processes. As such, it may be possible to exceed this 0.4~ML coverage by starting from a substantially thicker $\rm{B_2O_3}$ layer and relying on thermal desorption to remove any excess \cite{sear2017germanium}, or by depositing directly onto bare \cite{sear2017germanium,sear2018thermal} or oxidised (100) diamond \cite{o2013diamond,o2015extremely} followed by annealing at lower temperatures.This termination method also opens up possible pathways to boron $\delta$-layer growth, given the recent success of Kuntumalla~\textit{et al.} in forming nitrogen-doped $\delta$-layers in diamond by a process of surface termination and subsequent CVD overgrowth \cite{kuntumalla2021nitrogen}. It is interesting to consider the boron concentration which might be achievable if further developments allow for the oxygen to be removed \cite{korobtsov2006b2o3}, or alternative boron sources employed, and CVD then used to incorporate the boron into the diamond. Assuming a boron surface concentration of 0.4~ML, as has been achieved here, and similar growth dynamics to the nitrogen case - namely, an average 10\% retention rate during overgrowth and a layer thickness $\leq3$~nm - a layer grown using this approach would have a boron concentration above $2\times10^{20}$~cm\textsuperscript{-3}, close to the metallic and superconducting doping regime in diamond \cite{yokoya2005origin}. At the present time the best boron-doped delta layers in diamond that have been achieved, formed using methods employing CVD growth in bespoke ``delta doping'' CVD reactors \cite{butler2017nanometric} employing rapid gas switching systems \cite{balmer2013transport,vikharev2016novel} and typically subsequent etching or polishing \cite{fiori2012situ}, have a comparable doping density ($\sim5 - 8\times10^{20}$~cm\textsuperscript{-3}) while requiring more complex processing and instrumentation to achieve. As such, this new termination process may be an initial step in developing a simpler, surface-based approach to forming boron-doped delta layers in diamond, which with process optimisation may improve on the doping density and layer thinness achievable with existing approaches. Such an outcome may enable realisation of the improvements in quantum transport that are predicted for boron-doped $\delta$-layers in diamond \cite{chicot2014electronic} and the scalable fabrication of nanoscale boron-doped diamond systems and devices.  

\section{Conclusion}
In summary, we have demonstrated the ability to form a boron-oxide termination on (100) diamond, starting from an initially hydrogen-terminated diamond surface and using $\rm{B_2O_3}$ as a molecular precursor. Surface sensitive XPS and NEXAFS indicate that the $\rm{B_2O_3}$ interacts minimally with the hydrogen-terminated diamond surface, while annealing to 950\celsius\, to desorb the hydrogen causes a reaction between the $\rm{B_2O_3}$ adlayer and the termination-free (100) surface. From our XPS analysis we determine that carbon-boron single bonds form at the surface, and that the surface-bound boron has oxygen bonded to it, with the majority of the boron-oxide termination coming in the form of $\rm{B=O}$ moieties bound to the surface. Additionally, NEXAFS measurements show angle-dependent intensity variation, indicating a high degree of orientational ordering of the resulting chemistry. Combining our observations from these two forms of spectroscopy suggests that our termination process has yielded a highly homogeneous termination, with some minority of surface sites - likely defect and step edge sites - terminated with alternative boron-oxide configurations. This boron-oxide termination process may be adaptable to other boron species, enabling the formation of diamond/boron nitride interfaces, and may be a step in a path to a non-CVD approach to boron incorporation in diamond, thereby enabling alternative pathways to fabricating doped delta layers in diamond.

\section*{Acknowledgements}
This research was undertaken using the Soft X-ray Spectroscopy beamline at the Australian Synchrotron, part of ANSTO. This work was performed in part at the Melbourne Centre for Nanofabrication (MCN) in the Victorian Node of the Australian National Fabrication Facility (ANFF). This work was supported by the Australian Research Council under the Discovery Early Career Researcher Award (DECRA) scheme (DE190100336).

\section*{References}
\bibliography{C_B_bib}

\end{document}


\title[Formation of a boron-oxide termination for the (100) diamond surface]{Supplementary Information:\\Formation of a boron-oxide termination for the (100) diamond surface}

\author{Alex K. Schenk$^{1}$, Rebecca Griffin$^{1}$, Anton Tadich$^{1,2}$, Daniel Roberts$^{3}$, Alastair Stacey$^{3,4}$}

\address{$^{1}$ Department of Mathematical and Physical Sciences, School of Computing, Engineering and Mathematical Sciences, La Trobe University, Victoria 3086, Australia}
\address{$^{2}$ Australian Nuclear Science and Technology Organisation (ANSTO), Clayton, VIC 3168, Australia}
\address{$^{3}$ School of Science, RMIT University, Melbourne, Vic 3001, Australia}
\address{$^{4}$  Princeton Plasma Physics Laboratory, Princeton University, 100 Stellarator Rd, Princeton, NJ, 08540, USA}

\ead{A.Schenk@latrobe.edu.au}

\vspace{10pt}

%
\section*{Coverage Calculations}
Adlayer coverages have been calculated using the ratio of the B1s and C1s core level intensity, as acquired with a photon energy of 400~eV. This consistent choice of photon energy removes the need to compensate for different photon fluxes, and the core level kinetic energies – $E_K\sim175$~eV for B1s, $E_K\sim115$~eV for C1s – ensure that the measurement is highly surface sensitive. 

The ratio of the experimentally acquired B1s and C1s intensities is compared to a calibration curve calculated using the following approach, which we have employed in other works \cite{schenk2015formation,sear2017germanium}. We have started by calculating the intensity produced by each individual layer and then summing over all layers, appropriately attenuated, producing the total intensity of the C1s signal $I_{C1s}^U$ in the absence of a molecular layer, using the expression:

\begin{equation}
I_{C1s}^U = \sum_{N=0}^{N=\infty} n_C\times \sigma_{C1s}^{400} \times T(E_{K,C1s}) \times \exp\left( -N\times \frac{t_D}{\lambda_S(E_{K,C1s})\cos\theta}\right) 
\end{equation}
where:
\begin{itemize}
	\item $n_C$ is the atomic layer density of diamond (100) (2 atoms per unit cell, or $1.5747\times10^{15}$~cm\textsuperscript{-2}). 
	\item $\sigma_{C1s}^{400}$ is the photoabsorption cross-section of the C1s core level for a photon energy of 400~eV \cite{yeh1985atomic}.
	\item $T(E_{K,C1s})$ is the transmission function of the hemispherical analyser at the measurement conditions used for the kinetic energy corresponding to the C1s core level.
	\item $t_D$ is the atomic layer thickness in (100) diamond (0.895~\AA).
	\item $\lambda_S(E_{K,C1s})$ is the photoelectron mean free path in diamond at the kinetic energy corresponding to the C1s core level (8.33~\AA) as calculated using the TPP-2M method \cite{tanuma1994calculations}.
	\item $\theta$ is the photoelectron emission angle with reference to the surface normal ($\theta = 0\degree$).
	\item $N$ is a layer index.
\end{itemize}

We then account for the attenuation of this intensity by an overlayer with coverage $\Theta$, using an expression which accommodates the fact that a partial layer will attenuate the signal from some areas of the surface more than others. Specifically, the C1s intensity with attenuation accounted for $I_{C1s}$ is given by:

\begin{eqnarray}
	I_{C1s} 
	&= I_{C1s}^U \times \left(\left(\Theta - q\right)\exp\left(-\frac{(q+1)t_{B_2O_3}}{\lambda_{B_2O_3}(E_{K,C1s})\cos\theta}\right) \right.\\
	&\quad\left.+ \left(q+1-\Theta \right) \exp\left(-\frac{qt_{B_2O_3}}{\lambda_{B_2O_3}(E_{K,C1s})\cos\theta}\right)\right) \nonumber   
\end{eqnarray}

where
\begin{itemize}
	\item $t_{B_2O_3}$ is the atomic layer thickness of $B_2O_3$ (2.77~\AA).
	\item $\lambda_{B_2O_3}(E_{K,C1s}$ is the photoelectron mean free path in $B_2O_3$ at the kinetic energy corresponding to the C1s core level (6.45~\AA) as calculated using the TPP-2M method \cite{tanuma1994calculations}.
	\item $q$ is equal to the number of complete layers – for example, for a coverage $\Theta=2.3$~ML, $q=2$, and for a coverage $\Theta=0.6$~ML, $q=0$. 
\end{itemize}
The intensity of the B1s signal is calculated using a similar approach. The unmodified intensity for 1 ML of $B_2O_3$ is given by:
\begin{equation}
	I_{B1s}^U = n_B \times \sigma_{B1s}^{400}\times T(E_{K,B1s})
\end{equation}
where
\begin{itemize}
	\item $n_B$ is the boron density of one monolayer of $B_2O_3$ ($5.31\times10^{14}$~cm\textsuperscript{-2}). 
	\item $\sigma_{B1s}^{400}$ is the photoabsorption cross-section of the B1s core level for a photon energy of 400~eV \cite{yeh1985atomic}.
	\item $T(E_{K,B1s})$ is the transmission function of the hemispherical analyser at the measurement conditions used for the kinetic energy corresponding to the B1s core level.
\end{itemize}
For submonolayer coverages $(\Theta \leq 1)$, the intensity $I_{B1s}$ is assumed to be:
\begin{equation}
	I_{B1s} = \Theta \times I_{B1s}^U
\end{equation}
For coverages greater than 1 ML $(\Theta > 1)$, it is necessary to account for self-attenuation within the layer:
\begin{eqnarray}
		I_{B1s} 
		&= I_{B1s}^U \left( (q+1-\Theta) \sum_{N=0}^{N=q-1} \exp\left(-\frac{Nt_{B_2O_3}}{\lambda_{B_2O_3}(E_{K,B1s})\cos\theta}\right) \right.\\ 
		&\quad\left.+ (\Theta - q) \sum_{N=0}^{N=q} \exp\left(-\frac{Nt_{B_2O_3}}{\lambda_{B_2O_3}(E_{K,B1s})\cos\theta}\right) \right) \nonumber
\end{eqnarray}

where
\begin{itemize}
	\item $\lambda_{B_2O_3}(E_{K,B1s})$ is the photoelectron mean free path in $B_2O_3$ at the kinetic energy corresponding to the B1s core level (9.28~\AA) as calculated using the TPP-2M method \cite{tanuma1994calculations}.
\end{itemize}
This expression splits the intensity into two terms. To demonstrate by example, if the molecular coverage is assumed to be 2.3 ML, then the first term accounts for those areas of the surface where there are 2 layers (70\% of the surface), while the second term accounts for those areas of the surface where there are 3 layers (30\% of the surface). This layer-by-layer summation approach – both for the substrate signal and the adlayer - is preferred in this case over other simpler expressions for coverage calculation – which often use integration - as it explicitly accounts for fact that different regions of the surface will experience different amounts of signal attenuation which, for measurements in the surface sensitive regime, can be significant. 

Following these calculations, the calculated signals are combined into a ratio $I_{B1s}/I_{C1s}$, and plotted as a function of coverage, allowing the experimentally measured intensity to provide a measure of coverage.
 
For the case of the post-anneal boron-oxide terminated surface, the same approach has been used, assuming the same mean free path as B2O3, although the boron layer density for 1 ML has been assumed to be equal to the diamond surface carbon density.

\section*{References}
\bibliography{Supp_C_B_bib}